\documentclass[prl,twocolumn,floatfix,amsfonts]{revtex4}
\usepackage{graphicx,graphics,color,epsfig}
\usepackage{bm}
\usepackage{amsmath}
\usepackage{amssymb}
\begin{document}
\preprint{}
\title{Orthogonality Catastrophe in Bose-Einstein Condensates}
\author{Jun Sun, Olen Rambow, and Qimiao Si}
\affiliation{Department of Physics \& Astronomy, Rice University,
Houston, TX 77005--1892}

\begin{abstract}
Orthogonality catastrophe in fermionic systems 
is well known: in the thermodynamic limit, the overlap 
between the ground state wavefunctions with and without
a single local scattering potential approaches zero algebraically
as a function of the particle number $N$. Here we examine
the analogous problem for bosonic systems. In the homogeneous case,
we find that ideal bosons display an orthogonality stronger 
than algebraic: the wavefunction overlap behaves as 
${\rm exp}[-\lambda N^{1/3}]$ in three dimensions and as 
${\rm exp}[-\lambda N/\ln ^2 N]$ in two dimensions.
With interactions, the overlap becomes finite but is still
(stretched-)exponentially small for weak interactions.
We also consider the cases with a harmonic trap,
reaching similar (though not identical) conclusions.
Finally, we comment on the implications of our results for 
spectroscopic experiments and for (de)coherence phenomena.
\end{abstract}
\maketitle


In the area of correlated electrons, there is a long history
of studying quantum impurity problems such as the Kondo 
effect~\cite{Hewson}. A broad range of phenomena arise depending
on the type of impurity, the nature of the bulk electron system,
and the way they are coupled. The conceptual basis for this
richness is provided by the orthogonality 
catastrophe~\cite{Anderson}. It deals
with the effect of a single scattering potential on the many-body
states of an ideal $N$-fermion system. 
The many-body ground state wavefunction is a Slater determinant
of the $N$ lowest single particle wavefunctions. 
Since the impurity potential affects each and every one 
of the single-particle states, its effect on the many-body
state is significantly amplified: in the thermodynamic limit
the ground state wavefunction in the presence of the impurity
potential ($|{\rm gs}'>$) is orthogonal to its counterpart
in the absence of the impurity potential ($|{\rm gs}>$).
For large but finite $N$, the wavefunction overlap,
$S \equiv  |< {\rm gs} | {\rm gs}'>|$, depends on $N$
in an algebraic form:
\begin{eqnarray}
S_{\rm fermion} \sim N^{-\alpha} .
\label{s-fermion}
\end{eqnarray}
The exponent $\alpha = \delta^2 / d \pi^2$ 
when the scattering contains only 
an s-wave component; here,
$\delta$ is the $s-$wave 
scattering phase shift of the single-particle state 
at the Fermi energy,
and $d$ the dimensionality.
The initial work of Anderson provided 
an upper bound. An exact solution was later achieved 
in the work of Nozi\`eres 
and collaborators~\cite{Nozieres}. 
For an illuminating discussion, including the connection with
the Friedel sum rule, see Ref. ~\cite{Hopfield}.

Cold atoms provide a setting to engineer many-boson systems with 
a variety of quantum phases~\cite{Greiner}.
It is natural to ask what happens
when local defects are introduced into these systems.
Here we consider the orthogonality effect
in a Bose-Einstein condensate (BEC). For ideal homogeneous bosons,
the problem will be solved exactly in a very simple way.
For the cases with interactions and/or confining background potentials,
we will carry out analyses perturbatively in the impurity potential.

{\it Ideal bosons in a uniform background:~~} 
In this case, the effect of an impurity can be determined exactly.
The ground state condensate wavefunction is 
\begin{eqnarray}
|{\rm gs}> = \varphi_0({\bf x}_1)
\varphi_0({\bf x}_2) ...
\varphi_0({\bf x}_N) ,
\label{gs-bec}
\end{eqnarray}
where $\varphi_0$ is the single-particle ground state wavefunction.
$|{\rm gs}'>$ has the same form, with $\varphi_0$ being replaced by
$\varphi_0'$, the single-particle ground state wavefunction in the 
presence of the impurity. The overlap, $S$, is then simply
\begin{eqnarray}
S &=& s^N 
\label{S=sN}\\
s &\equiv& <\varphi_0|\varphi_0'>
\label{s=phiphiprime}
\end{eqnarray}

Consider first the case of three dimensions. 
We choose a spherical box of radius $R$, and a fixed boundary 
condition $\varphi (r=R) =0$. The impurity is placed 
at the center:
\begin{eqnarray}
\Delta H=V\Theta(a-r)
\label{DeltaH-def}
\end{eqnarray}
where $a$ is its size.
Without a loss of generality, the potential will be taken 
as repulsive. we define $\psi$ as the radial part 
of the single-particle wavefunction multiplied by $r$.
In the region $0<r<a$, 
${\psi_I} =A \sin ({k_I} r)$, with ${k_I}^2=\frac{2m}{{\hbar}^2}
(\epsilon-V)$. In the region $a<r<R$, ${\psi_{II}}
=B \sin({k_{II}} r+\delta)$, with ${k_{II}}^2 =\frac{2m}{{\hbar}^2}\epsilon$,
where $\delta$ is the phase shift.
The fixed boundary condition 
at $r=R$
yields $k_{II}=\frac{\pi-\delta}{R}$. 
The continuity of 
$\psi$
and its derivative at
$r=a$ gives rise to
\begin{equation}
\frac{\tan ({k_I} a)}{k_I}=\frac{\tan ({k_{II}} a+\delta)}{k_{II}}
\label{continuity-3D}
\end{equation}

\begin{figure}[h]
\centerline{\psfig{file=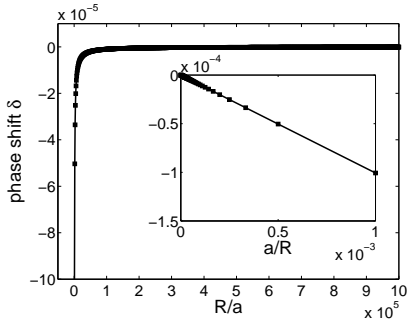,height=4.3cm}}
\caption{%
Phase shift as a function of the system size in three
dimensions, with $\bar{V}=0.1$. 
Inset: the phase shift vs. $a/R$
 }
\label{fig:phase-shift-3d}
\end{figure}

The phase shift for the 
single-particle ground state
in the thermodynamic limit ($R/a \rightarrow \infty$)
is found to be
\begin{equation}
\delta_{3d}=-\frac{a}{R}f(V)
\label{delta-3d}
\end{equation}
where 
$f(V)= \pi (1-\frac{\tanh(\bar V)}{\bar V})$, with ${\bar
V}\equiv (\frac{2mVa^2}{{\hbar}^2})^{1/2}$.
The $a/R$ dependence reflects the fact that 
the single particle state sits at the edge of the band,
where the density of states is proportional to $\sqrt{\epsilon} 
\propto a/R$. The numerical result [Fig.~\ref{fig:phase-shift-3d}]
shows the validity of this form already for 
$R/a$ of the order $10^3$.

From the forms of the wavefunctions in the presence and in the 
absence of the impurity, we determine the overlap of the 
single-particle wavefunctions
\begin{eqnarray}
s = 1 - {1 \over 6} (1+ { 3 \over {4 \pi^2}}) \delta_{3d}^2 ,
\label{s-3d}
\end{eqnarray}
where we have used the fact that $\delta_{3d}$ is always small
in the thermodynamic limit.
Combining Eqs.~(\ref{S=sN},\ref{s-3d},\ref{delta-3d}), we find 
a stretched-exponential form for the overlap of the many-body
ground states:
\begin{eqnarray}
S_{\rm 3d} = {\rm exp}[-\lambda (V) N^{1/3}]
\label{S-boson-3D}
\end{eqnarray}
where $\lambda (V) = \frac{1}{6}(1+\frac{3}{4\pi^2})a^2
f^2(V)\left (\frac{4\pi n_0}{3}\right )^{\frac{2}{3}}$, with $n_0$ being the
particle density.

Consider next the case of two dimensions -- a disk of radius $R$.
The 
ground state 
wavefunction $\psi$
is now a linear
combination of the zeroth order Bessel and Neumann functions:
$\psi_I =A J_0(k_I r)$ and $\psi_{II}=B [J_0(k_{II}r)+\tan(\delta) 
N_0(k_{II}r)]$.
The fixed boundary condition at $r=R$
now gives
$k_{II}=\frac{\frac{3\pi}{4}+\delta}{R}$. The continuity at $r=a$
of
$\psi$ and its derivative leads to
\begin{equation}
\frac{k_I J_1(k_I a)}{J_0(k_I a)}=\frac{k_{II}
[J_1(k_{II}a)+\tan(\delta)N_1(k_{II}a)]}{J_0(k_{II}
a)+\tan(\delta)N_0(k_{II} a)}
\end{equation}
Using the limiting forms of the Bessel and Neumann functions
appropriate for $a/R \rightarrow 0$,
we find the following phase shift of the single-particle ground 
state
\begin{eqnarray}
\delta_{\rm 2d}=\frac{\pi}{2}\frac{1}{\ln(R/a)} .
\label{delta-2d}
\end{eqnarray}
Note that it is independent of the potential strength.
The finite value of the density of states at the band edge would
have implied a phase shift
that is independent of $R/a$. 
The exact result, on the other hand, contains a logarithmic
correction factor. 
In Fig.~2, we plot the numerical solution
which confirms the logarithmic factor.

\begin{figure}[h]
\centerline{\psfig{file=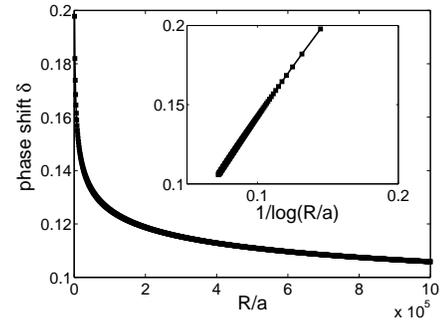,height=4.3cm}}
\caption{%
The analogous plot for two dimensions, with
$\bar{V}=1$. Inset: phase shift vs. $1/\ln(R/a)$.
 }
\label{fig:phase-shift-2d}
\end{figure}
Taking these results together we find the overlap of the many-body
ground states in two dimensions:
\begin{eqnarray}
S_{\rm 2d} = {\rm exp}[-\alpha N/\ln^2 N] ,
\label{S-boson-2D}
\end{eqnarray}
where $\alpha = \frac{\pi^2}{2}\left [
\frac{\int B^2(x)}{\int
J_0^2(\frac{3\pi}{4}x)}-5\left (
\frac{\int J_0(\frac{3\pi}{4}x)B(x)}{\int
{J_0}^2 (\frac{3\pi}{4}x)} \right )^2 \right ]$, with
$B(x)=N_0(\frac{3\pi}{4}x)-J_1(\frac{3\pi}{4}x)x$ and $\int
f(x)\equiv \int_0^1 f(x)x dx$.

So far, the results are exact. In order to generalize to situations 
in which exact solutions are not readily available (see below),
we now turn to a perturbative treatment of the impurity potential.
To the second order in a linked cluster expansion,
the overlap between the many-body ground state
wavefunctions is~\cite{Mahan}:
\begin{eqnarray}
S &=& {\rm exp}[-C] \nonumber \\
C &=& \frac{1}{2} \sum_{n \ne {\rm gs}} {
{| <n|\Delta H | {\rm gs}> |^2 }
\over 
{(E_n -  E_{\rm gs})^2 }
} .
\label{S-through-C}
\end{eqnarray}
The impurity scattering can be rewritten as 
$\Delta H = V a^\dagger({\bf x}={\bf 0}) a({\bf x}={\bf 0})$,
where $a^\dagger({\bf x})$ and $a({\bf x})$ are the boson field
operators.
The off-diagonal long-range order of a BEC
implies that we can set $a_0$ and $a_0^{\dagger}$ 
to be $\sqrt{N}$. The impurity Hamiltonian becomes
$\Delta H = \Delta H_1 + 
V/L^d (N + \sum_{{\bf p} \ne {\bf 0} \ne {\bf p'}} 
a_{\bf p}^{\dagger} a_{\bf p'} )$,
where the part of concern to us is
\begin{eqnarray}
\Delta H_1 = V {\sqrt{N} \over L^d} \sum_{{\bf p} \ne {\bf 0}}
(a_{\bf p}^{\dagger} + a_{\bf p})
\label{Delta-H-1}
\end{eqnarray}
Here, $L$ is the linear dimension of the system.
Combining Eqs.~(\ref{S-through-C},\ref{Delta-H-1}),
and using the dispersion relation $\epsilon({\bf p}) = p^2/2m$,
we have
\begin{eqnarray}
C = \frac{V^2}{2} {N \over L^d} 
\frac{(2m)^2}
{(2\pi \hbar)^d}
\int_{\hbar\pi/L}^{\infty} d^d {\bf p} ~p^{-4} .
\label{C-perturbative}
\end{eqnarray}
In three dimensions, $C$ has a divergence of the form
$L \propto N^{1/3}$, which is the same as the exact
result [Eq.~(\ref{S-boson-3D})]. 
(For the spherical box geometry considered earlier,
the perturbative result is found to fully agree with
what comes out of Eq.~(\ref{S-boson-3D}) when $\lambda(V)$
is expanded to order $V^2$.)
In two dimensions, 
the divergence becomes $L^2 \propto N$. Compared with the exact
result [Eq.~\ref{S-boson-2D}], the perturbative result 
recovers the $N$ factor in the exponential but misses
the multiplicative logarithmic correction.

The perturbative treatment also provides the physical picture 
for our results.
To see this, we rewrite the expression for $C$ in terms of the
dissipation spectral function, $Im\chi_0^{-1} (\omega)$, as follows,
\begin{eqnarray}
C &=& \frac{1}{2}
\int_{(E_1-E_{\rm gs})} 
^{\infty} d \omega 
{ { {\rm Im} \chi_0^{-1} (\omega) }
\over
\omega^2} 
\label{C-through-chi0}
\\
{\rm Im }
{\chi_0^{-1} (\omega) }
&= &\sum_{n \ne {\rm gs}} | <n|\Delta H | {\rm gs}> |^2 \times 
\nonumber\\
&&\times \delta (\omega -E_n + E_{\rm gs})
\label{chi0}
\end{eqnarray}
where $E_1$ is the energy of the first excited state with a non-zero
$<n|\Delta H | {\rm gs}>$.
The form of the impurity potential, Eq.~(\ref{Delta-H-1})
implies that the dissipation spectral function is simply
proportional to the single-particle density of states:
\begin{eqnarray}
{\rm Im} \chi_0^{-1} (\omega>0)\propto \omega^{{d \over 2} -1} .
\label{chi01-ideal-bosons}
\end{eqnarray}
The exponent $({d \over 2} -1)$ reflects the quadratic nature
of the dispersion of the low-lying excitations.
It is less than $1$ for both three and two dimensions; 
in the terminology adopted in the dissipated two-level system
literature~\cite{Leggett}, both are sub-ohmic~\cite{Kirchner}. 
This abundance of low-lying excitations is responsible
for the strong orthogonality in a BEC.

{\it Weakly interacting bosons in a uniform background:~~} 
Unlike for fermions, even weak interaction is a relevant perturbation 
for bosons. We use the standard 
Bogoliubov transformation~\cite{Bogoliubov,Lifshitz},
\begin{eqnarray}
a_{\bf p} = {1 \over \sqrt{1 - L_{\bf p}^2}}
[b_{\bf p} + L_{\bf p} b_{-{\bf p}}^{\dagger}] .
\label{bogoliubov}
\end{eqnarray}
Under this transformation, the quadratic part of the Hamiltonian becomes
\begin{eqnarray}
H_0 = \sum_{{\bf p} \ne {\bf 0}} \xi({\bf p}) b_{\bf p}^{\dagger} b_{\bf p}.
\label{bogoliubov-h-0}
\end{eqnarray}

For $p \gg 2mu$ (where 
$u = \sqrt{U n_0/m}$ is the sound velocity, with
$U$ being the effective contact interaction amplitude),
$L_{\bf p}$ nearly vanishes and we recover the 
non-interacting limit, including 
Eq.~(\ref{Delta-H-1})
and $\xi ({\bf p}) \approx \epsilon 
({\bf p})$.
For $p \ll 2mu $, on the other hand, 
$L_{\bf p} \approx -1+p/mu$,
and it follows that 
\begin{eqnarray}
\xi ({\bf p}) &=& u p \nonumber \\
\Delta H_1 &=& 
V {\sqrt {N} \over L^d}
\sum_{{\bf p} \ne {\bf 0}}{}^{\prime}
\sqrt{{p}\over {2mu}} (b_{\bf p}^{\dagger} +b_{\bf p}) ,
\label{Delta-H-1-interacting}
\end{eqnarray}
where the prime denotes that the summation is up to about
$2mu$.
Eq.~(\ref{C-perturbative}) is then replaced by
\begin{eqnarray}
C && \approx {V^2\over {2}}
{N \over L^d} 
{1 \over (2\pi\hbar)^d} (2m)^2 \times \nonumber \\
&&\times
\left [ \frac{1}{(2 mu)^3}
\int_{\hbar \pi/L}^{2mu} d^d {\bf p}~ p^{-1}
+
\int_{2mu}^{\infty} d^d {\bf p} ~p^{-4}
\right ] .
\label{C-interacting}
\end{eqnarray}
It is convergent in both three and two
dimensions, implying a finite wavefunction overlap.
The resulting overlap in the thermodynamic limit
depends on the interaction $U$ 
in the following 
(stretched-)exponential forms:
\begin{eqnarray}
S_{\rm U,3d} = {\rm exp}[-\lambda'/\sqrt{U}] \nonumber \\
S_{\rm U,2d} = {\rm exp}[-\alpha'/U] 
\label{S-boson-U}
\end{eqnarray}
where $\lambda' \approx \frac{V^2\sqrt{n_0}m^{3/2}}
{4\pi^2\hbar^3}$ and 
$\alpha' \approx \frac{3V^2m} {8\pi\hbar^2}$.

This conclusion can also be seen through the 
form of the dissipative-bath spectral function,
which, at low-energies, now takes the form
[{\it cf.} Eqs.~(\ref{chi0},\ref{Delta-H-1-interacting})]
\begin{eqnarray}
{\rm Im} \chi_0^{-1} (\omega>0)\propto \omega^{d} .
\label{chi01-interacting-bosons}
\end{eqnarray}
Its super-ohmic nature in two and three dimensions
implies that $C$ [{\it cf.} Eq.~(\ref{C-through-chi0})]
is infrared convergent~\cite{Leggett}.

{\it Ideal bosons in a harmonic confining potential:~~} 
Consider an isotropic harmonic trap with frequency $\Omega_0$.
The thermodynamic limit is defined by keeping 
\begin{eqnarray}
N \Omega_0^d = {\rm const.}
\label{h.o.-thermodynamic-limit}
\end{eqnarray}
as both $N$ and $1/\Omega_0$ go to infinity~\cite{Dalfovo}.
(For instance, it ensures a finite energy per particle for ideal
fermions.)

We consider an impurity located at the center of the trap:
$\Delta H = V a^{\dagger} ({\bf x =0}) a ({\bf x =0})$.
Using $a({\bf x}) = \sum_{\bf n} \varphi_{\bf n} ({\bf x}) a_{\bf n}$,
where ${\bf n} \equiv (n_1,\cdot \cdot \cdot,n_d)$, and 
the single-particle eigenfunctions 
\begin{eqnarray}
\varphi_{\bf n} ({\bf x}) &=& 
\left( { {m\Omega_0} \over {2 \hbar}} \right )^{d/4}
\Pi_{i=1}^d (n_i!)^{-{1 \over 2}}\left ( 
{ {m\Omega_0} \over {2 \hbar}} \right )^{{n_i \over 2}}
\times \nonumber \\
&& \times \left ( x_i - {\hbar \over 
{m\Omega_0}}{d \over {d x_i}} \right )^{n_i}
{\rm e}^{-m\Omega_0 r^2 / 2\hbar}  ,
\label{phi-ho}
\end{eqnarray}
we write the linear part of the impurity Hamiltonian as
\begin{eqnarray}
\Delta H_1 &=& 
V \sqrt { {N} {\Omega_0}^{d}}
\sum_{{\bf n} \ne {\bf 0}} A_{2{\bf n}} (a_{2{\bf n}}^{\dagger} + 
a_{2{\bf n}}) .
\label{Delta-H-1-ho}
\end{eqnarray}
Here, we have defined  
\begin{eqnarray}
A_{2{\bf n}} \equiv 
\frac{\varphi_{2{\bf n}}^*(0)\varphi_{\bf 0}(0)}
{\Omega_0^{d / 2}}
=
\frac{(-1)^{\sum_{i=1}^d n_i}}
{(\pi \hbar /m)^{d/2}}
\prod_{i=1}^{d}\left [
{{(2n_i-1)!!}\over{(2n_i)!!}} \right ]^{{1}\over{2}} .
\label{A-n}
\end{eqnarray}
From Eq.~(\ref{S-through-C}),
we have
\begin{eqnarray}
C=\frac{V^2}{2}
(N\Omega_0^{d}){1 \over (\hbar \Omega_0)^{2}}
\sum_{{\bf n} \neq {\bf 0}} 
{A_{2{\bf n}}^2 \over (2n)^2} ,
\label{C-ho-ideal}
\end{eqnarray}
where $n=\sum_{i=1}^{d} n_i$.
At large $n$, we find that $\sum_{n_1,
\cdot \cdot \cdot,n_d} \delta (n-\sum_{i=1}^d n_i)
A_{2{\bf n}}^2$ is proportional to $\sqrt{n}$
in three dimensions and approaches a constant in two
dimensions~\cite{sterling}, so 
the summation over $n$ in Eq.~(\ref{C-ho-ideal})
is convergent at ultraviolet.
$C$ is then proportional to $\Omega_0^{-2}$, which, using the thermodynamic
limit, is equivalent to $N^{2/d}$. The overlap between the ground state
wavefunctions is then
\begin{eqnarray}
S_{\rm ho,3d} &\propto& {\rm exp}[-\lambda_{ho} N^{2/3}] \nonumber
\label{C-ho-3D}\\
S_{\rm ho,2d} &\propto& {\rm exp}[-\alpha_{ho} N ] 
\label{C-ho-2D}
\end{eqnarray}
where $\lambda_{ho}= \frac{V^2}{4\sqrt{\pi}}\zeta(\frac{3}{2})
(m/\pi\hbar^2)^3(N\hbar^3\Omega_0)^{\frac{1}{3}}$ and
$\alpha_{ho}=V^2(m/\hbar^2)^2/48$.

{\it Weakly-interacting bosons in a harmonic confining potential:~~} 
To understand the effect of interactions, we first note on one important
consequence of the thermodynamic limit, Eq.~(\ref{h.o.-thermodynamic-limit}).
In the non-interacting case, it is seen, from the ground state wavefunction
$\varphi_{\bf 0}({\bf x}) = \left( { {m\Omega_0} / {\pi \hbar}} 
\right )^{d/4} {\rm e}^{-m\Omega_0 r^2 / 2\hbar}$, that 
the central density of the condensate, $n({\bf x}={\bf 0}) \equiv N 
|\varphi_0({\bf x}={\bf 0})|^2$, is of the order of $\sqrt{N}$
in the thermodynamic limit.

In the interacting case, on the other hand, 
$n({\bf x}={\bf 0})$ is well-known to be of order unity 
in the thermodynamic limit~\cite{Dalfovo}.
In this limit,
the interaction term dominates over the kinetic term~\cite{Baym}.
It follows from the Gross-Pitaevskii equation
that $n({\bf x}={\bf 0}) = \mu / U$ where 
the chemical potential $\mu$ is finite in the thermodynamic limit.
This implies that $\varphi_0({\bf x}={\bf 0})$ in the interacting case
is a factor of 
$N^{-1/4}$ smaller than its counterpart in the non-interacting case.
The wavefunctions of the low-lying excited 
states should contain a similar reduction factor.
On the other hand, the energies of the collective modes 
remain linear in $\Omega_0$. 
It follows [{\it cf.}
Eqs.~(\ref{S-through-C},\ref{Delta-H-1-ho},\ref{A-n})]
that interactions weaken the orthogonality effect, as in the
uniform case.
The precise form of the overlap depends on the details 
of the excited-state wavefunctions, which can be determined
from the Bogoliubov-de Gennes equations in a harmonic
potential; this will be discussed elsewhere.

{\it Experimental implications:~~} In addition to the theoretical
significance, the orthogonality effect may also be directly
probed in experiments. One way is to perform the analog 
of the x-ray edge measurement in metals~\cite{Hopfield,Mahan}.
Consider a condensate co-existing with a separate species of
atoms that are considerably more dilute and are localized (by a deep
optical potential well that only these atoms see).
The photo-absorption or luminescence corresponding to
a transition between two levels of this second species
of atoms would ordinarily be a sharp delta function. 
(In practice, the spectral width of a hyperfine transition 
for atoms in a BEC can be as narrow as $\sim$100 Hz 
or even $\sim$10 Hz~\cite{Hulet}.)
However, the two atomic levels will provide different scattering 
potentials for the atoms of the condensate. The weight of the
delta function -- which is precisely the overlap of the condensate
wavefunctions corresponding to these two different potentials 
-- would then have to vanish in the thermodynamic limit due to the 
orthogonality catastrophe. A finite spectral weight can arise 
only when the condensate atoms go to the excited states under
the new potential, where the excitation energy serves as a cutoff
for the infrared divergence.
This results in a one-sided spectrum~\cite{Hopfield}, which
can be divergent or vanishing at the edge depending on the 
degree of orthogonality for the low-lying excited states.

The orthogonality may also be manifested in the 
time evolution of a condensate after a sudden introduction
of a local potential. The orthogonality makes it rather hard
for the system to evolve into the new ground state.
In other words, the density distribution will tend to keep
its initial profile; the impurity is ``hardly visible'' to the 
condensate.

Yet another implication is on the coherence and decoherence phenomena.
Consider, for instance, localized atoms in a BEC. An effective Kondo
problem arises when both the localized atoms as well as 
the itinerant atoms of the condensate contain real or pseudo-
spin degrees of freedom~\cite{Zwerger,Cirac}.
Strong orthogonality makes the spin flip process entirely
incoherent.


To summarize, we have studied the orthogonality effect in 
Bose-Einstein condensates. For ideal bosons, the overlap 
of the ground state wavefunctions when a single local 
scattering potential changes from one value to another
vanishes in a stretched-exponential form in
the thermodynamic limit. With interactions,
the overlap becomes finite but is small for weak
interactions; its dependence on the interaction strength is 
typically stretched-exponential as well.
These effects can be probed using spectroscopic experiments
in cold atoms, which can be tuned from being essentially
ideal to strongly interacting. The effects also have significant
implications for the coherence and decoherence phenomena in
bosonic systems.


We would like to thank E. Abrahams, P. W. Anderson, 
K. Damle, R. Hulet, H. Pu and, in particular, C. M. Varma
for useful discussions. 
The work has been supported by NSF Grant No.\ DMR-0090071
and the Robert A. Welch Foundation.


\begin{thebibliography}{99}
\bibitem{sterling} It is important to use the strong form
of Sterling's formula, $n! \approx \sqrt{2\pi n} n^n {\rm e}^{-n}$,
which implies 
$(2n-1)!!/(2n)!! \approx 1/\sqrt{\pi n}$.
\end{thebibliography}
\end{document}